\documentclass[sigconf, nonacm]{acmart}

\newcommand\holodoi{10.61981/ZFSH2309}
\newcommand\holopages{38-44}
\newcommand\holovolume{1}
\newcommand\holoissue{1}
\newcommand\holoyear{2023}
\newcommand\holoauthors{\authors}
\newcommand\holotitle{\shorttitle} 
\newcommand\holoavailabilityurl{}
\newcommand\holopagestyle{plain} 

\usepackage{subcaption}

\begin{document}
\title{A Conceptual Model of Intelligent Multimedia Data Rendered using Flying Light Specks}

%%
%% The "author" command and its associated commands are used to define the authors and their affiliations.
\author{Nima Yazdani}
\affiliation{%
  \institution{University of Southern California}
  \city{Los Angeles}
  \state{CA}
  \country{USA}
}
\email{nimayazd@usc.edu}

\author{Hamed Alimohammadzadeh}
\affiliation{%
  \institution{University of Southern California}
  \city{Los Angeles}
  \state{CA}
  \country{USA}
}
\email{halimoha@usc.edu}

\author{Shahram Ghandeharizadeh}
\affiliation{%
  \institution{University of Southern California}
  \city{Los Angeles}
  \state{CA}
  \country{USA}
}
\email{shahram@usc.edu}

%%
%% The abstract is a short summary of the work to be presented in the
%% article.
\begin{abstract}
A Flying Light Speck, FLS, is a miniature sized drone configured with light sources to illuminate 3D multimedia objects in a fixed volume, an FLS display.  A swarm of FLSs may provide haptic interactions by exerting force back at a user's touch.  This paper presents a conceptual model for the multimedia data to enable content-based queries.  The model empowers users of an FLS display to annotate the illuminations by adding semantics to the data, extending a multimedia repository with information and knowledge.  We present a core conceptual model and demonstrate its extensions for two diverse applications, authoring tools with entertainment and MRI scans with healthcare.
\end{abstract}

\maketitle

%%% do not modify the following Holodeck block %%
%%% Holodeck block start %%%
\pagestyle{\holopagestyle}
\begingroup\small\noindent\raggedright\textbf{Holodecks Reference Format:}\\
\holoauthors. \holotitle. Holodecks, \holovolume(\holoissue): \holopages, \holoyear.\\
\href{https://doi.org/\holodoi}{doi:\holodoi}
\endgroup
\begingroup
\renewcommand\thefootnote{}\footnote{\noindent
This work is licensed under the Creative Commons BY-NC-ND 4.0 International License. Visit \url{https://creativecommons.org/licenses/by-nc-nd/4.0/} to view a copy of this license. For any use beyond those covered by this license, obtain permission by emailing \href{mailto:info@holodecks.quest}{info@holodecks.quest}. Copyright is held by the owner/author(s). Publication rights licensed to the Holodecks Foundation. \\
\raggedright Proceedings of the Holodecks Foundation, Vol. \holovolume, No. \holoissue.\\
%\ %
%ISSN 2150-8097. \\
\href{https://doi.org/\holodoi}{doi:\holodoi} \\
}\addtocounter{footnote}{-1}\endgroup
%%% Holodecks block end %%%

%%% do not modify the following Holodecks block %%
%%% Holodecks block start %%%
\ifdefempty{\holoavailabilityurl}{}{
\vspace{.3cm}
\begingroup\small\noindent\raggedright\textbf{Holodecks Artifact Availability:}\\
The source code, data, and/or other artifacts have been made available at \url{\holoavailabilityurl}.
\endgroup
}
%%% Holodecks block end %%%

\section{Introduction}
\label{sec:intro}

There is an abundance of 3D data and devices that produce this data.
They pertain to diverse applications ranging from healthcare to entertainment.
For example, with healthcare, MRI scanners produce 3D scans of a patient's brain.
With entertainment, a graphic artist may use a 3D authoring tool such as Maya or Blender to produce animated sequences.
Flying Light Specks~\cite{shahram2021,shahram2022,mmsys2023}, FLSs, enable 3D displays to render this data without sacrificing either its meta-data or the raw device data.
%3D displays are relatively new with the introduction of Flying Light Specks, FLSs~\cite{shahram2021,shahram2022,mmsys2023}.
%An FLS is a miniature sized drone configured with light sources, processing, storage and networking capabilities.
%Swarms of cooperating FLSs will illuminate complex 2D and 3D shaped in a fixed volume, a 3D display~\cite{dv2023}.
%FLSs will provide both tactile (skin-based) and kinesthetic (force) haptic feedback, enabling a user to experience the shape and stiffness of an illumination. 
A conceptual challenge is how these displays will integrate into the eco-system of today's devices to illuminate their 3D data and facilitate advanced interactions.
This paper presents a conceptual data model to address this challenge.
Its objective is to articulate the requirements at a high level to engage all stake holders, both technical and non-technical.
The model will facilitate future logical (mathematical) and physical (software and hardware) models.
The proposed data model is extensible, enabling a designer to extend it with new concepts specific to an application.

\begin{figure}
  \centering
  \includegraphics[width=\linewidth]{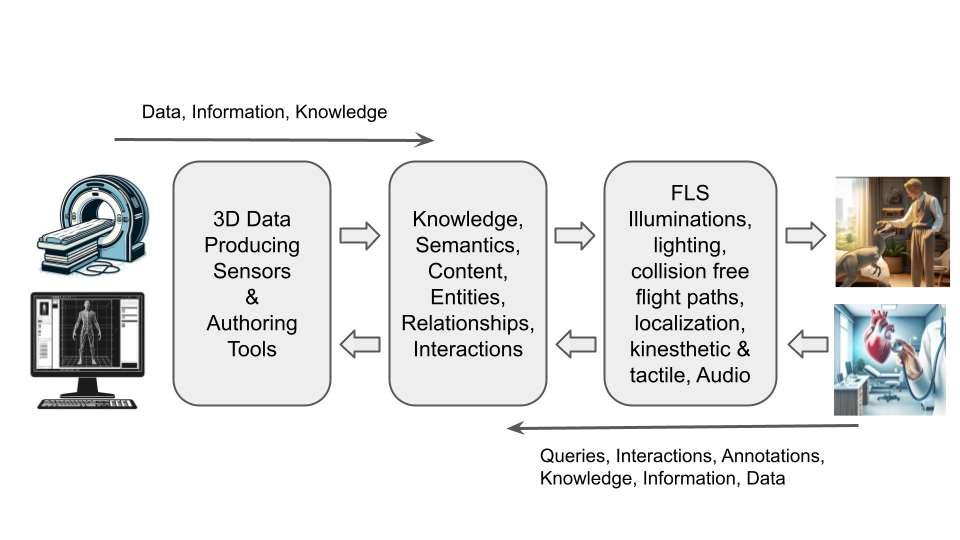}
  \caption{An eco-system for devices and authoring tools, and 3D FLS displays.}
  \label{fig:DataInfoKnow}
\end{figure}

Figure~\ref{fig:DataInfoKnow} provides an end-to-end overview of a framework that includes the proposed conceptual data model.
At one end of the spectrum, devices and authoring tools produce raw data with diverse metadata.
This metadata may be rich in content.
For example, the metadata authored by a graphic artist using a 3D authoring tool may pertain to settings of algorithms that perform complex tasks, e.g., keyframing for interpolations that produce animated sequences~\cite{surveyInterpolation2018}.  
This data is abstracted and transformed into information, a summarization that is more interpretable by humans, e.g., MRI scan of the patient's brain shows dark matter.
Finally, data and information is transformed into knowledge that provides an understanding of the content of the physical and virtual worlds, e.g., the patient has seizures because of head injury.

At the other end of the spectrum,
%shown in Figure~\ref{fig:DataInfoKnow}, 
a 3D FLS display provides its users with illuminations and diverse forms of interactions including haptics.
Users may query the system for the relevant data, information, and knowledge.
In addition, users may annotate the illuminations to contribute additional data, information, and knowledge to the system. 

%The ER model that will be discussed in this paper is designed to proficiently manage those same layers of data, information, and knowledge associated with 3D visualizations created using Flying Light Specks. This ER model will serve as a foundational framework to enable more interactive and efficient engagement with FLS-based visualizations, offering significant advantages in multimedia applications.

In this paper, we present a conceptual model that describes the end-to-end spectrum of Figure~\ref{fig:DataInfoKnow}.
The immediate benefit of the conceptual data model is two folds.  First, it facilitates design and implementation of efficient algorithms for the FLS display.
Second, it enables content-based queries and annotations that contribute knowledge, realizing intelligent multimedia data. %rendered using FLS displays.
Consider each in turn.
An algorithm that enables a 3D FLS display may use the semantics of multimedia data.
To illustrate, consider an animated sequence produced by an authoring tool.  
To illuminate this sequence, an FLS display must compute the flight path of FLSs as a function of time.
This can be done in several ways.
In~\cite{shahram2022}, the animated sequence is reduced to a sequence of mesh files that are subsequently processed using the Motill~\cite{shahram2022} algorithm.
While highly parallelizable, Motill is a resource intensive algorithm that detects motion across the points of different mesh files.
A more efficient algorithm may use the proposed conceptual model that identifies the key frames that have motion.  
This may be the basis of a new algorithms with a significantly lower complexity than Motill.
Similarly, with content based queries, we may use the information and knowledge readily available from devices and authoring tools to provide relevant results.
This will enable language models that understand and respond to domain specific queries.

The {\bf contributions} of this paper include:
\begin{enumerate}
\item A core conceptual data model for intelligent multimedia data displayed using an FLS display.
Section~\ref{sec:core}.
\item Extensions of the core to MRI healthcare data and authoring tools for animation.
Sections~\ref{sec:animationER} and~\ref{sec:mirER}.
\end{enumerate}
Section~\ref{sec:related} presents related work.  Brief conclusions and future research directions are presented in Section~\ref{sec:future}.

\section{Related Work}\label{sec:related}

The Flying Light Speck, FLS, as a miniature sized drone configured with light sources, processing, storage and networking capabilities to illuminate complex multimedia shapes was first introduced in~\cite{shahram2021}.
3D FLS displays~\cite{shahram2021,shahram2022,shahram2022b,dv2023} are in the same class of systems that materialize virtual objects in a fixed volume.
These include fast 3D printing~\cite{t1000}, Claytronics as physical artifacts using programmable matter consisting of catoms~\cite{matter2005}, Roboxels as cellular robots that dynamically configure themselves into the desired shape and size~\cite{roboxel1993}, 
BitDrones~\cite{gomes2016bitdrones} and GridDrones~\cite{griddrones2018} as interactive nano-drones.

Motill~\cite{shahram2022} is an algorithm that computes the flight path of FLSs for an animated sequence consisting of mesh files displayed at a pre-specified rate, e.g., 24 mesh files per second.
Motill is resource intensive and time consuming.  For an animated sequence that is displayed repeatedly, it is more efficient to store Motill's computed flight paths in a file once and read the file for each repeated display. A file format to store these flight paths is described in~\cite{mmsys2023}.
This study presents a conceptual data model for its file format. 
The conceptual model presented here is different because it captures information and knowledge.
It strives to maintain semantics and facilitate content-based queries.

Conceptual models are recognized as an important tool in several computing disciplines including databases, multimedia, and software engineering.
The Entity-Relationship (ER) data model used in our study dates back to the seminal work of Chen~\cite{chen75} and instructed by numerous computer science textbooks, e.g.,~\cite{erchapter}.
Conceptual models and the idea of data transformed to information and knowledge is present in diverse scientific applications including Earth Sciences~\cite{nasa2002,nasa2005}.

\section{Core Conceptual Data Model}\label{sec:core}

%First, let's construct a foundational abstract ER diagram that can be tailored to our subsequent applications in both entertainment and medicine.
%This section introduces a foundational conceptual data model that is subsequently tailored for two diverse applications of FLS displays, entertainment and health care. 

The core data model strives to provide a two-way interaction between the user and the display, enriching the user's overall experience.
Each FLS is a computer with the capability to execute algorithms that understand and interpret the data illuminated by its swarm.
One may conceptualize a swarm as one or more server racks of today's data centers that facilitates rich interactions with the user.
Backed by a conceptual data model, the FLS display becomes more than a one-way channel for visual and haptic rendering.
It enables its users to inquire about the materialized objects, annotate and manipulate them, and provide summaries.
%It evolves into an interactive platform that not only illuminates 3D objects but also allows users to inquire about data, visualize the data, have haptic interactions with the data, annotate and manipulate the displayed data.
Through these interactions, users may contribute additional data, information, and knowledge to the multimedia repository.
See Figure~\ref{fig:DataInfoKnow}.

%{\bf The reason the FLS display needs a conceptual data model in the first place is because, as it exists, the FLS display lacks the capability to understand or interpret the data it presents. The integration of an Entity-Relationship model addresses this limitation. With the ER model, the FLS display becomes more than a one-way channel for visual information. It evolves into an interactive platform that not only presents 3D objects but also allows users to inquire about and manipulate the displayed data. This completely changes what it means to consume media, creating a two-way interaction between the user and the FLS display, enriching the overall experience.}

The resulting conceptual data model consists of entity sets (rectangles) and relationship sets (diamonds). Both may have attributes, represented as ovals. Multiple entity sets may be involved in a relationship set.
The relationship set is represented as a diamond.
A single line from an entity set to a relationship set suggests an entity may exist without participating in a relationship. For instance, the connection between the Object and Make Noise indicates that an object may exist without acoustics. Conversely, a double line signifies mandatory participation, meaning an entity can only exist if it participates in that relationship. For example, the linkage between Flight Paths and Objects dictates that an entity within the Objects entity set must establish a relationship with Flight Paths\footnote{This is justified because we are modelling data for an FLS display and the flight path of FLSs to illuminate the objects.}. One may use an aggregated entity set (dotted rectangle) to define a relationship as an entity that may participate- in another relationship. %This aggregated entity set may participate in a relationship with another entity set. 
An entity set may also participate recursively with itself through a relationship set as shown between the Objects and Consists-Of relationship sets.

\begin{figure}
  \centering
  \includegraphics[width=\linewidth]{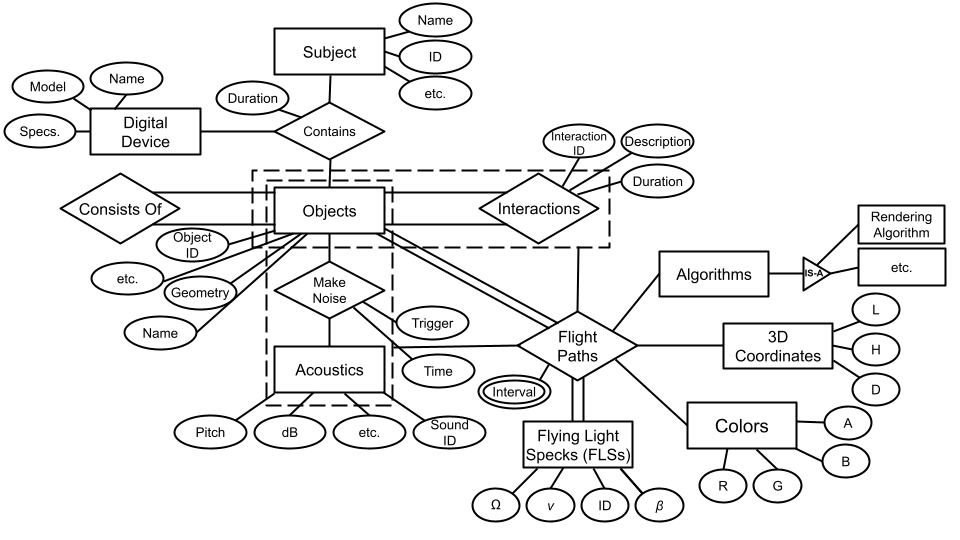}
  \caption{An extensible core conceptual data model.}
  \label{fig:abstractER}
\end{figure}

%\subsection{Abstract Model Structure}
%Figure~\ref{fig:abstractER} consists of eight entity sets and five relationship sets. It shows 
Figure ~\ref{fig:abstractER} shows a Subject entity set that contains Objects. This relationship has a Duration attribute that specifies when an object enters and exits the subject's realm relative to the start of the subject. The Objects entity set contains the attributes Object ID, Name, Geometry, etc. The attributes of the Objects entity set are also connected to the Acoustics entity set by the Make Noise relationship set. The Make Noise relationship set includes two key attributes: Time and Trigger.
These serve as indicators for the timing and initiation of sound emissions from an object. The Time attribute specifies the exact moment in a visual sequence when an object's acoustic element should be activated. 
%On the other hand, 
The Trigger attribute identifies the specific type of user-object interaction that should initiate the sound. For instance, if a user pets a virtual cat or pats a virtual dog, the Trigger would initiate the respective sounds of a purr or a bark. The Acoustics Entity set stores the attributes of each noise an object makes with the attributes: Sound ID, pitch, dB, Frequency, and any other attributes needed to achieve the full depth of the audio. This relationship between the Objects entity set and the Acoustics entity set is aggregated as one entity set denoted with a dashed rectangle that participates in the relationship set Flight Paths denoted with a single line.

The Objects entity set also has a recursive relationship through the Consists-Of and the Interactions relationship sets. The Consists-Of relationship illustrates how each object in the Objects entity set can be composed of smaller objects, which in turn may be made up of even smaller objects, and so forth. The Interactions set illustrates a recursive relationship in which an object may interact with another object, which in turn may interact with yet another, and so on. The Interactions relationship set has the attributes: Interaction ID, Duration, Description, etc. This relationship between the Objects entity set and the Interactions is aggregated as its own entity set denoted with a dashed rectangle that participates in Flight Paths.
This many-to-many relationship set requires the total participation of the Objects and the FLSs entity sets. The FLSs entity set outlines the velocity model of an FLS ($\nu$), the duration of its flight time on a fully charged battery ($\beta$), the amount of force needed by the FLS to simulate a certain level of resistance (\( N \)), and the battery's charging time ($\Omega$)\cite{shahram2022}. ID is the unique identifier of an FLS. The relationship between Objects, Interactions, and Flight Paths is a fundamental aspect of this model. Each object generates its own Flight Paths, which are then adjusted based on the interactions among the objects. Similarly, when users modify these Flight Paths, it directly influences the Objects and Interactions Entity Sets. This process ensures that the model adapts to user modifications, creating a more tailored and effective system.
%We identify an FLS uniquely using its ID.  %Since many FLSs will likely have attributes in common, an ID is assigned to distinguish each one. 

The Flight Paths relationship set engages either the 3D Coordinates and Colors entity sets, the Algorithms entity set, or both. This flexibility accommodates various scenarios: the digital device may supply data that directly populates the 3D Coordinates and Colors sets; alternatively, the Algorithms entity set alone may suffice to calculate the coordinates and colors of the FLSs. In some scenarios, both may work in conjunction to generate comprehensive flight paths. 

The 3D Coordinates entity set specifies the Length $L$, Height $H$, and Depth $D$ positioning for an FLS. The Colors entity set determines the brightness levels of various lights that the FLS is tasked to display. Figure~\ref{fig:abstractER} employs the Red, Green, Blue, and Alpha (RGBA) color scheme. The Algorithms entity set stores the algorithms necessary for computing the flight paths of each FLS. As each FLS has its own computational capabilities, this entity set utilizes the Objects, Interactions, FLSs, and other potential entity sets as inputs to generate each individual flight path on-board. The Algorithms entity set also branches, through the Is-A relationship, into a subclass, specifically for Rendering algorithms, which handles the dynamic color changes for each FLS. This Algorithms set is designed to be extendable, capable of incorporating a wide range of algorithms used for FLS manipulation. These algorithms are responsible for the calculations necessary for the bidirectional interactions between users and the display. By incorporating algorithms for lighting properties, stiffness, or the general behavior of different states of matter, the FLS flight paths can adapt to the inputs received by users when they feel, bend, or poke displayed objects for example. Depending on the specific application of our ER model, the Algorithms entity set must be specialized to include functions tailored to the task at hand. These specialized functions could continue to use the Objects, Interactions, FLSs, and other entity sets as inputs, adapting to any changes in their attributes.

Flight Paths also has an Interval attribute which describes how long an FLS renders a color at a 3D coordinate. The Interval attribute is a multi-valued (double ellipsoid) because an FLS may be required to render the same color and 3D coordinate at different times. An example is a hummingbird beating its wings where the same positions and colors may be required for multiple flaps in place.

The ER diagram of Figure~\ref{fig:abstractER} consists of two parts, one that describes the content of an application and a second that describes how to display the content using an FLS display.
The next two sections adapt this foundational abstract ER diagram to suit our two diverse applications: entertainment and healthcare.

% \begin{definition}An FLS flight path Consists-Of at least four attributes and two optional attributes: the 3D coordinates visited by the FLS, the color(s) rendered at a coordinate, an interval for this rendering, and the label of the object being rendered. Depending on whether or not an object has any interaction with other elements in the subject or if it produces any acoustics, there will also be an interaction and acoustic attribute participating as well.\end{definition}

\section{Entertainment: 3D Authoring Tools}\label{sec:animationER}

To integrate this ER model with existing animation authoring tools like Maya~\cite{derakhshani2012introducing}, Blender~\cite{villar2021learning}, ZBrush~\cite{keller2011introducing}, and others, we observe that these tools follow a similar progression from data to knowledge. Our ER model facilitates this progression, culminating in an FLS display. In traditional animation workflows, graphic artists rely on a keyboard, a mouse, and one or more 2D monitors to create 3D animations. These authoring tools export files that capture a wealth of metadata about the animated scene. Our Entity-Relationship (ER) model can take this rich data to generate flight paths for the FLS display, elevating the viewer's experience to an immersive 3D realm. But the innovation doesn't stop at viewing; the FLS display, backed by the ER model, transforms into an interactive animation studio. Graphic artists may manipulate the 3D objects in the scene directly, effectively turning the FLS display into both a viewing and an authoring tool. This opens up new dimensions of creativity, for example, allowing graphic artists to sculpt FLS-animated "clay" into complex figures, while also labeling objects, interactions, and extending them with acoustics. The result is a more dynamic, interactive, and enriched animation process.

%\subsection{Current Standard of Animation}
%\subsection{Today's State of The Art}
%In the realm of 3D animation, 3D authoring tools such as Blender have set the industry standard for creating life-like animations through keyframing techniques. 

\begin{figure*}
  \centering
  \begin{subfigure}[t]{0.48\linewidth}
    \centering
    \includegraphics[width=\linewidth]{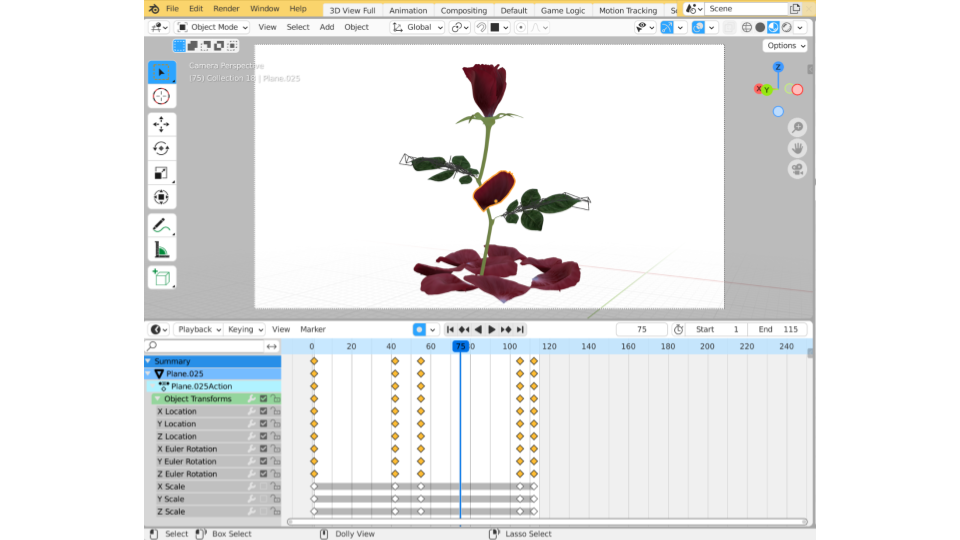}
    \caption{Keyframing.}
    \label{fig:KeyFraming}
  \end{subfigure}
  \hfill
  \begin{subfigure}[t]{0.48\linewidth}
    \centering
    \includegraphics[width=\linewidth]{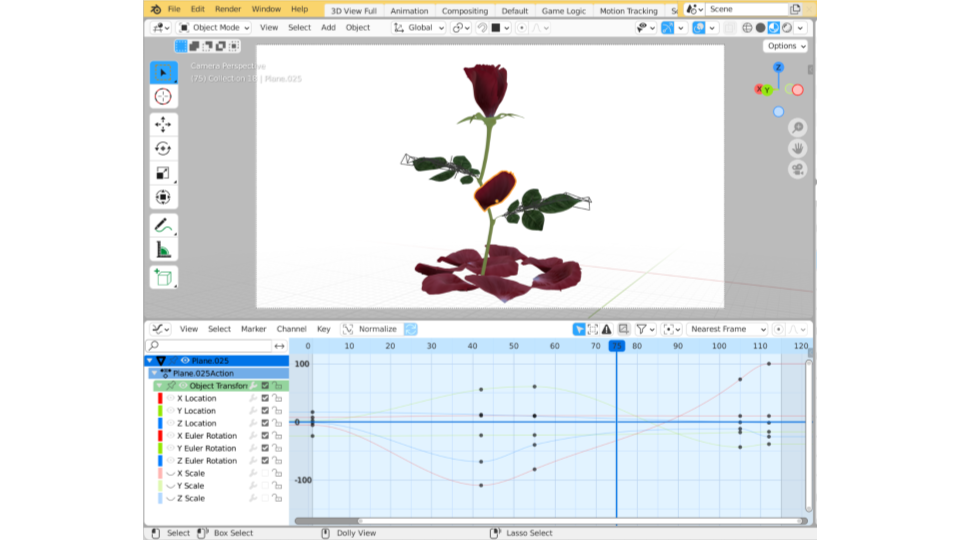}
    \caption{Bezier Interpolations.}
    \label{fig:BezierCurves}
  \end{subfigure}
  \caption{Blender.}
  \label{fig:blender}
\end{figure*}

\noindent{\bf Today's practices:}
Keyframing is a technique to create life-like animations expeditiously using interpolation algorithms.
The idea is for a graphic artist to specify a sequence of key frames while the algorithms compute the transitions that change one frame to the next.
This process starts with a graphic artist
%The process begins by 
defining keyframes at critical points along a timeline, which dictate the position and orientation of the object being animated. For example, with the rose petal animation, depicted in Figure~\ref{fig:KeyFraming}, the graphic artist defines the first keyframe that set the petal's starting position and the last keyframe that places it on the ground. Additional keyframes can be inserted in-between to introduce nuances like twists and curves in the petal's falling path. In the bottom pane of Figure~\ref{fig:BezierCurves}, these 5 keyframes are represented by the 5 vertical lines of diamonds.

%\begin{figure}
%  \begin{center}
    %\includegraphics[width=0.9\columnwidth]{figs/KeyFraming.png}
  %\end{center}
%  \caption{Keyframing in Blender}\label{fig:KeyFraming}
%\end{figure}

%\begin{figure}
%  \begin{center}
    %\includegraphics[width=0.9\columnwidth]{figs/BezierCurves.png}
  %\end{center}
  %\caption{Bezier Interpolations in Blender}\label{fig:BezierCurves}
%\end{figure}
An authoring software such as Blender may offer a graphic artist a choice of interpolation algorithms 
%The software also allows for different types of keyframe interpolation—
such as 
e.g., linear or Bezier, to govern how animated properties change over time. 
%Bezier interpolation, in particular, is favored for its ability to produce smooth and natural movements. These Bezier interpolation functions are represented by the thin lines at the bottom of Figure~\ref{fig:BezierCurves}, connecting each of the five keyframes (vertical constelations of points). 
%The bottom pane of Figure~\ref{fig:BezierCurves} shows the Bezier interpolation functions as the thin lines connecting each of the five keyframes denoted as vertical constellation of points.
Once the keyframes and interpolation types are set, the software calculates the values of animated properties, like position and rotation, over the specified time.

% {\bf In the next paragraph, is there a reference to back the claim that MP4 is the most commonly used export format?}

% The most commonly used export format for these Blender generated animations is MP4, prized for its wide compatibility and efficient balance between quality and file size. However, the MP4 format comes with inherent limitations, particularly when it comes to the granularity of information it retains. 
% %The compression process involved in creating an MP4 file 
% MP4 effectively "flattens" all the individual elements of a scene into a single, inseparable sequence of I, P, and B frames~\cite{mpeg}, a video stream. This means that the rich, object-specific information, so crucial during the animation phase, becomes inaccessible for querying or further manipulation once the file is exported.  {\bf What about MP7?}

\subsection{Advantages of a Conceptual Data Model for Authoring Tools}

The Algorithms entity set of Figure~\ref{fig:abstractER} describes the keyframe-based approach.
A swarm of FLSs corresponding to the changing portion of a keyframe may execute the algorithm to compute their flight paths.
%This current keyframe-based approach can be captured by our conceptual data model, where each object contains keyframes that are animated using the corresponding algorithms and eventually converted into flight paths for the FLS display. 
Each object in the FLS display retains its full set of individualized attributes, interactions, and compositional elements, allowing users to query these objects.  % with unparalleled accuracy. 
For example, a user may issue the following query, ``Who is hiding in the closet?" and the system may accurately respond and even use the FLS display to illuminate the hidden entity.

The proposed conceptual model aligns more naturally with human cognitive processes, particularly our innate spatial awareness. Humans naturally perceive the world in three dimensions, and our ability to understand spatial relationships is finely tuned. Animating 3D objects in a 3D space simulated with FLSs taps into this natural aptitude, making it easier and more intuitive to understand complex scenes. This also allows multiple graphic artists to visualize the same display from different angles and simultaneously edit and mold the objects within it. Any changes made are captured and updated in the conceptual model, offering a dynamic, interactive, and adaptable animation environment. In this collaborative setting, character designers, graphic artists, sound designers, and other creators may collaborate in real-time within the same 3D space, enhancing the creative process and allowing for immediate feedback and iterative design. 

%Repetitious
%By allowing for accurate real-time querying, retaining object-specific information, and displaying 3D content in a 3D space, %the FLS and ER diagram approach to animation 
%our conceptual model
%opens up new vistas in the field of animation, offering a more interactive and intuitive experience.

\subsection{Extending the Core Model for Animation}
\begin{figure}
  \centering
  \includegraphics[width=\linewidth]{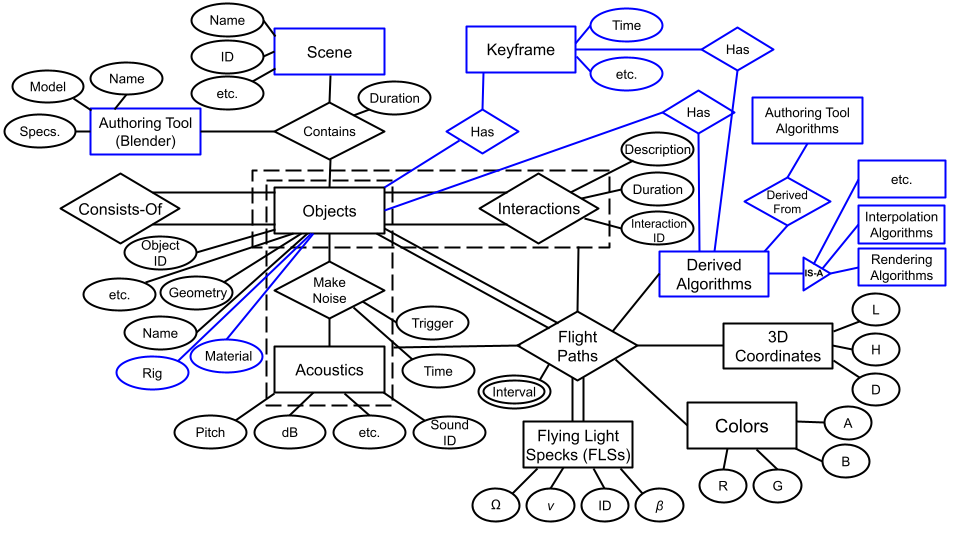}
  \caption{ER Model for Animation Tools}
  \label{fig:ER Model for Animation Tools}
\end{figure}

%In order to achieve the advantages of our system we must also adapt our conceptual framework for storing data generated by authoring tools and executing queries of that data, we need to customize the ER model depicted in Figure~\ref{fig:abstractER} to specifically address Application 1. In this adapted model shown in Figure~\ref{fig:ER Model for Animation Tools}, the Subject entity set transforms into the Scene entity set, now featuring attributes such as Scene Name, Scene ID, Media ID (indicating the clip to which the animation belongs), Description, Author Name, and any other key identifiers essential for both viewing and editing the animation.
We use an ontology to tailor the core model of Figure~\ref{fig:abstractER} to generate the extended ER diagram of Figure~\ref{fig:ER Model for Animation Tools}.
The Subject entity set is renamed to the Scene entity set, maintaining attributes Name and ID, and can be expanded to include any other key identifiers deemed essential for both viewing and editing the animation.

Similarly, the Digital Device entity set is named the Authoring Tool entity set, representing software platforms like Autodesk Maya, Blender, Zbrush among others. This updated entity set provides an understanding of the tools used to create animations and includes attributes like Name, Model, and software specifications.

The Objects entity set is enhanced, now with attributes: Object ID, Name, Material, Rig (the "skeletal" structure of an animated object), and Geometry.

% The Consists Of relationship set now includes a pre-computed attribute called Hidden, which determines whether a sub-object is visually represented within a larger object. For instance, when displaying a bag, its contents may not need to be visible. However, if the bag is made of transparent plastic, the contents may need to be shown on the FLS display. The computation of this attribute is inspired by Blender's ray-tracing algorithms, which simulate the behavior of light interacting with transparent surfaces. These algorithms trace rays of light from the camera through each pixel in the image plane and into the scene, accounting for factors such as refraction, reflection, transparency, and absorption to accurately represent the object.

The entity set originally named Algorithms has been renamed to Derived Algorithms. This set is derived via a new relationship set called Derived From, which links it to a new Authoring Tool Algorithms entity set. The latter contains algorithms employed by the 3D animation authoring software for its own simulations, which serve as the foundation for the Derived Algorithms now used to calculate the flight paths.

The Derived Algorithms entity set also branches, through the Is-A relationship, into subclasses, specifically for Rendering and Interpolation algorithms. The Rendering Functions handle the dynamic color changes for each FLS, while the Interpolation Functions facilitate the seamless transition between keyframes whenever a property of an object is changing over time. The Derived Algorithms entity set is designed for flexibility and scalability, enabling the integration of a diverse array of animation algorithms. An illustrative example is Blender's sophisticated ray tracing algorithms, which handle optical phenomena like refraction, reflection, transparency, and absorption. This entity set can incorporate similar ray tracing algorithms, reimagined for 3D displays as opposed to Blender's 2D framework. Additionally, it may encompass a range of general functions that model various physical states of matter, providing a comprehensive toolkit for accurately simulating real-world physics in animation.

The Objects entity set now branches through a new Has relationship set to a newly introduced Keyframe entity set. The Keyframe entity set includes a Time attribute and is designed to be extendable, for example, to allow for the addition of further attributes to specify which properties of the object are undergoing changes in a particular keyframe. The Time attribute marks the specific moment the keyframe occurs in the animation. Furthermore, the Keyframe entity set is linked to the Derived Algorithms set through another Has relationship set, specifying the interpolation algorithm used for that particular keyframe. Similarly, the Objects entity set maintains a Has relationship with the Derived Algorithms set to access the required Rendering Algorithms for dynamically coloring the object.

The remaining foundational elements of the original ER model remain intact, ensuring that the adapted model is both robust and tailored to the specific queries needed for animation. % needs of animation application queries. 
%This modified ER diagram may readily incorporate data from animations created in Blender. 
In this model, each object in a scene is an entity in the Objects entity set, complete with its geometry, material, geometry, rig, and keyframes for any property changes. For instance, to depict a position change in a rose petal, multiple keyframes would be required and stored in the Keyframes Entity set and realted through the Has relationship set to the rose petal's entry in the Objects entity set. The Interactions and Consists Of entities can be populated using an interpretation of the 3D rendering with an algorithm from the Derived Algorithms set, or can be manually annotated by the viewer or graphic artist. The Algorithms entity set would be populated with all the functions needed to take the geometry, material, rig, and keyframes of an object, combined with the objects it consists of, its interactions, and the sounds it makes, to derive a flight path for every FLS emulating that object. 

\section{Healthcare: MRI Scans}\label{sec:mirER}

Magnetic Resonance Imaging, MRI, scanners create 3D images of the anatomical structure of the human body.
These images are used for both research and diagnosis of an illness or disorder.
There are two types of scans, structural and functional MRI (fMRI).
A structural scan is a single 3D image represented as a matrix of voxels.
A voxel corresponds to measurements from a small volume of body, quantifying a property such as the volume's (tissue) density or water content.
fMRI is a sequence of structural images as a function of time, describing the change in the tissue.
A popular use of fMRI is to analyze brain activity overtime and in response to a stimulant.

An immersive FLS display that illuminates either a structural or an fMRI scan may be the size of a room.
A physician may enter this room with an organ illuminated in front of them.
They may step (zoom) into the organ by having it expand around them to analyze individual voxels visually.
The physician may touch the voxels to experience their stiffness (this stiffness would be derived from the properties observed in the MRI cross referencing an external database that stores the necessary data to simulate an accurate relative stiffness and density).
The physician may shrink the rendering as appropriate to orient themselves relative to an organ before zooming (stepping) into it to examine its voxels.
The physician may annotate an organ, a voxel, or groups of voxels.
The physician may color code voxels to highlight a disease, and associate knowledge (e.g., identify a spreading cancer) with different organs, tissue and one or more voxels.

%The integration of the proposed Entity-Relationship model with an FLS display offers transformative potential for healthcare, particularly in the realm of medical imaging such as MRI scans, EEGs, and other similar technologies. Currently, the interaction with MRI scans is largely limited to observation, zooming, annotation, rotation, and parameter adjustment. While these interactions provide valuable insights, they lack the immersive and, tactile engagement that a 3D FLS display could offer. Virtual Reality (VR) and Augmented Reality (AR) technologies have attempted to bridge this gap, but they come with their own set of limitations.

A conceptual model for the structural and fMRI scans enables the aforementioned interactions with an FLS display. 
%The introduction of an ER model as the underlying architecture for the FLS display addresses these limitations. 
Acting as the 'brain' of the system, the model enables a more sophisticated interaction with the medical images. Physicians can go beyond basic observational interactions and engage with a true-to-life 3D representation of various anatomical structures. This enhanced interaction allows for a more comprehensive and nuanced analysis, facilitating the generation of in-depth diagnostic reports. Moreover, the FLS display, informed by the ER model, becomes an intelligent visualization tool that 'understands' what it is displaying. This not only enriches the experience for the physician but also provides an interface where health care providers can inquire about interactions shown and object's properties.

%\subsection{Current Practice of MRI Visualization}
\noindent{\bf Today's practices:}
Current MRI practices predominantly generate 2D images in single planes, often referred to as "slices." 
This 2D display format inherently limits the level of spatial comprehension, particularly for complex medical procedures, for example the removal of a Myxoma~\cite{abdumajidov2023giant}, an inner heart tumor. %For example, the removal of a Myxoma, current 2D MRI visualizations may not provide the most intuitive understanding of the intricate interactions between the tumor and the heart's chambers, which may be inhibiting blood flow.
Specifically the 2D framework causes a diminished depth perception in traditional MRIs. This limitation significantly hampers the ability to visualize complex structures such as white matter tracts within the nervous system, which are typically assessed using Diffusion Tensor Imaging (DTI), a specialized form of MRI.
Efforts to overcome the spatial awareness limitations inherent in 2D medical imaging have led to the exploration of augmented and virtual reality technologies. For example, in one study~\cite{duncan2023} researchers employed Virtual Reality (VR) head-mounted displays as a solution to enhance the visualization and labeling of lesions in brain scans. However, a primary drawback of this technology is VR-induced motion sickness, which can cause nausea in users after just 5 to 10 minutes of usage.

Another limitation of today's MRI visualization tools is the absence of tactile engagement. Currently, physicians rely on T1-weighted and T2-weighted MRI images to discern various tissue types and conditions. T1-weighted images visualize anatomical details and fat.
T2-weighted images are more sensitive to fluid and useful for detecting inflammation or edema.
Although physicians can adjust monitor settings to better interpret these grayscale images, they are limited to visual observation.
An FLS display will enable physicians to touch and poke the illuminated tissue to experience its density and stiffness directly.  
%This means they can infer tactile properties like tissue density or stiffness only indirectly, missing out on the valuable sensory information that actual physical interaction could provide.

%In addition to these limitations, 
At the time of this writing, the field of MRI imaging also faces inconsistencies in data presentation conventions. For instance, radiologists often view images with the brain's right side on the image's left, known as the "radiological convention," aligning with the view of the body from the foot of the bed. In contrast, neurologists prefer the "neurological convention," where the brain's left side matches the image's left. This inconsistency in brain imaging means that the interpretation of the X dimension is always a concern. Given the brain's left-right symmetry, discerning an image's orientation becomes challenging without context.
%Additionally, some MRI machines generate images in proprietary formats that require specialized software for viewing. This creates an additional barrier to the seamless sharing and interpretation of medical images. 

\subsection{Advantages of a Conceptual Model for MRI Data}

The current approach to displaying and interacting with medical images can be significantly enhanced by integrating a conceptual %data 
model with an FLS display. Such a display, enriched with detailed data, allows physicians to make more informed queries about the images. For instance, both radiologists and individuals with limited medical expertise could ask questions about constricted blood flow in the brain or the degree of inflammation in the heart since a patient's last MRI scan.

Moreover, this system enables tactile interaction with the 3D images. By incorporating data from CT scans—which also use voxels and are thus compatible with our conceptual model—we can accurately simulate the geometry, relative density, and other properties of each organ. This information is stored in the ER model and informs the flight paths and haptic capabilities of the FLSs, making kinesthetic interaction possible. For example, a doctor can brush his hand over a simulated brain and feel the changes in tissue to identify if a tumor is forming, and how the structural integrity of the organ will change if it is removed.

While AR and VR technologies offer a 3D experience, they come with limitations such as user nausea and motion sickness during extended use~\cite{duncan2023}. On the other hand, the FLS system materializes objects in a true 3D space, eliminating these issues. It undoubtedly provides a more comprehensive and easier-to-understand view, enabling physicians of various backgrounds and specialties to collaborate on a single patient's scans.

The field of MRI imaging is fraught with inconsistencies in data presentation, varying between radiologists and neurologists, among others. These inconsistencies can make it challenging to interpret MRI scans accurately and efficiently. Another advantage of our Entity-Relationship model is that it proposes a standardized approach to representing MRI data, aiming to unify these divergent practices and offer a consistent, 3D, easy-to-analyze format that can be universally adopted for MRI scans.

\subsection{Extending the Core Model for MRI Scans}

\begin{figure}
  \centering
  \includegraphics[width=\linewidth]{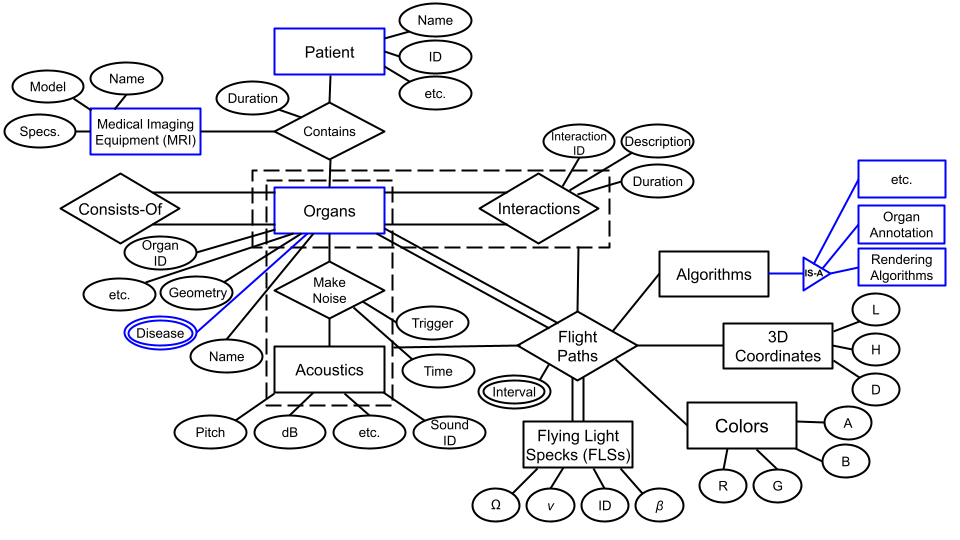}
  \caption{ER Model for MRI Data}
  \label{fig:ER Model for MRI Representation}
\end{figure}

%To extract the most utility from our abstract ER model in Figure~\ref{fig:abstractER} we must tailor it to store and display the data produced by MRI scans. 
Figure~\ref{fig:ER Model for MRI Representation} shows extensions of the core model of Figure~\ref{fig:abstractER} for MRI scans.
%In this adapted model 
%shown in Figure~\ref{fig:ER Model for MRI Representation} 
The Subject entity set is now renamed to the Patient entity set with the attributes Name, ID, and any other individual identifying information that is critical to a diagnosis. The Digital Device entity set will now be a Medical Imaging Equipment entity set, in this case an MRI entity set, with its attributes being Name, Model, and Specifications of the machine. The Objects Entity set is now adjusted to be the Organs Entity set with attributes: Organ ID, Name, Geometry, and Disease (the disease that each organ may be affected by). Disease is a double oval attribute denoting that it is a multidimensional attribute as one organ may be experiencing the symptoms of multiple diseases at once.
% reference images of ideal scans of the organ in question, and historical scans of the patients organs.

Our Algorithms Entity set has been extended to include Rendering and Organ Annotation Algorithms. The Rendering Algorithm is responsible for managing dynamic color changes, determining the colors of the FLSs in the Flight Paths set. Meanwhile, the Organ Annotation Algorithm processes data from MRI scans to analyze shape, size, location, signal intensity, texture, and more, thereby identifying and automatically labeling, areas of concern, tissues, muscles, and other organs within the MRI for targeted physician review. To accurately simulate tissue properties depicted in the MRI, the Algorithms Entity set also incorporates functions that model how organs interact with each other, including changes in interactions and alterations to their properties. For instance, if a surgeon wants to simulate the extraction of a brain tumor, the FLS must be capable of demonstrating the structural integrity of the surrounding tissue to indicate whether it could collapse. This is just one example; the algorithms required to simulate the properties and interactions of all bodily tissues are likely to be extensive.

The population of this Entity-Relationship diagram diverges from that of the authoring tool diagram. Unlike authoring tools, the MRI data primarily populates the 3D Coordinates and Colors Entity sets. These sets serve as inputs for the Flight Paths Relationship set and, crucially, the Algorithms Entity set. The Algorithms set, in turn, enriches the Organs Entity set along with its associated relationships and entities. Manual annotations by physicians can further augment the Objects, Interactions, and Consists-Of Entity sets, enhancing their accuracy and comprehensiveness. It's worth noting that the ER model can also be populated using data from other medical imaging techniques, such as CT scans, PET scans, and echocardiography for acoustic information. This allows for a more comprehensive and multi-modal approach to medical imaging.

\section{Conclusions and Future Research Directions} \label{sec:future}

This paper described a conceptual model for intelligent 3D multimedia data rendered using FLS displays.
The data is intelligent because it includes summaries in the form of information and knowledge, facilitating content-based retrieval.  A user may issue a query, illuminate its results using the FLS display, and annotate it with additional information and knowledge as necessary.

An immediate research direction is to develop a logical and physical counter-part of the conceptual model for Blender, a 3D authoring tool.
At its simplest form, the physical implementation enables Blender to output flight path of FLSs by computing coordinates and colors.
Its algorithms may use Motill~\cite{shahram2022} as their starting point and reduce its complexity by using the metadata (information and knowledge) provided by a graphic artist to Blender.
We will investigate the scalability and efficiency of this data model.
We will also investigate content-based queries described in Section~\ref{sec:animationER} and use of Blender's keyframing algorithms to compute FLS flight paths.

\section{Acknowledgments}
This research was supported in part by the NSF grant IIS-2232382.

%\clearpage

\bibliographystyle{ACM-Reference-Format}
\bibliography{sample}

\end{document}